\documentclass[12pt]{article}
\usepackage[latin1]{inputenc}
\usepackage[T1]{fontenc}
\usepackage[english]{babel}
\usepackage{amssymb}
\usepackage{amscd}
\usepackage{fancyhdr}
\usepackage{color}
\usepackage{epsfig}
\usepackage{graphicx}
\usepackage{bm}
\usepackage{subfigure}
\definecolor{red}{rgb}{1., 0., 0.}
\definecolor{blue}{rgb}{0., 0., 1.}
\definecolor{green}{rgb}{0.1, 0.7, 0.}
\definecolor{purp}{rgb}{1,0,1}
\newcommand{\bc}{\begin{cases}\begin{aligned}}
\newcommand{\ec}{\end{aligned}\end{cases}}

\newcommand{\eq}{\begin{equation}}
\newcommand{\fine}{\end{equation}}

\def\drawbox#1#2{\hrule height#2pt
        \hbox{\vrule width#2pt height#1pt \kern#1pt
              \vrule width#2pt}
              \hrule height#2pt}

\def\Asym#1#2{\vcenter{\vbox{\drawbox{#1}{#2}
              \kern-#2pt       
              \drawbox{#1}{#2}}}}

\newcommand {\beq} {\begin{equation}}
\newcommand {\eeq} {\end{equation}}
 \newcommand{\be}{\begin{eqnarray}}
\newcommand{\ee}{\end{eqnarray}}

\begin{document}

\begin{titlepage}

\begin{flushright}
\end{flushright}

\vspace {1cm}
\centerline{\Large \bf Getting ${\cal N}=1$ super Yang-Mills } 
\vspace {.4cm}
\centerline{\Large \bf from strings}

\vskip 1cm \centerline{\large Paolo Merlatti} \vskip 0.1cm

\vskip .5cm

\vskip 0.5cm \centerline{Departamento de Fisica de Particulas, Universidade de Santiago de Compostela} \centerline{E-15782, Santiago de Compostela, Spain}\centerline{ merlatti@fpaxp1.usc.es}

\vskip 1cm

\begin{abstract}
In the context of the gauge-string correspondence, we discuss the spontaneous partial breaking of supersymmetry. Starting from the orbifold of $S^5$, supersymmetry breaking leads us to consider the (resolved) conifold background and some of the gauge dynamics encoded in that geometry.
Using this gravity dual, we compute the low energy effective superpotential for such ${\cal N}=1$ theories. We are naturally led to extend the Veneziano-Yankielowicz one: glueball fields appear.   
\end{abstract}

\end{titlepage}
\tableofcontents

\section{Introduction}


Even if the most celebrated example of the gauge/string correspondence is the ${\cal N}=4$ AdS-CFT duality \cite{Maldacena:1997re} (see \cite{Aharony:1999ti} for a review), many interesting results have been also found for less (super)-symmetric theories (see \cite{Aharony:2002up,Imeroni:2003jk} for some extensive reviews).  

For the case of ${\cal N}=1$ super Yang-Mills theories in four dimensions ({\it{i.e.}} 4 real supercharges) the relevant background geometry is a warped product of four-dimensional Minkowski space times the conifold, in presence of background fluxes. There are two known explicit supergravity solutions corresponding to this case, the MN solution \cite{Maldacena:2000yy} and the KS one \cite{Klebanov:2000hb}. Both the solutions preserve the same amount of supersymmetry but they correspond to two different dual gauge theories, at least in the ultraviolet. It seems also that, despite describing similar infrared dynamics, they correspond to two different universality classes of confining theories. It is then important to understand how duality works case by case.  

 In this talk we make contact with a nice and fruitful way of looking at this duality. Such perspective has been introduced in \cite{Gopakumar:1998ki} for the topological string. There, it has been shown that in topological string theory the deformed conifold background, with $N$ 3-brane wrapped on $\mathbf{S}^3$ (equivalent to the $\mathbf{S}^3$ Chern-Simon theory \cite{Witten:1992fb}), is dual to the same topological string but without string modes and on a different background. The new background is the resolved conifold, whose parameters depend now on $N$.
Such gauge/geometry correspondence has been successfully embedded in superstring theory in \cite{Vafa:2000wi}. The central idea is the concept of geometric transition. The field theory has to be engineered via $D$-branes wrapped over certain cycles of a non-trivial Calabi-Yau geometry. The low energy dual arises from a geometric transition of the Calabi-Yau, where the branes have disappeared and have been replaced by fluxes. The simplest example \cite{Vafa:2000wi,Cachazo:2001jy} is pure ${\cal N}=1$ super Yang-Mills theory. Before the transition the field theory is engineered on $N$ $D5$ branes wrapped on the blown-up $\mathbf{ S}^2$ of a resolved conifold. After the transition we are instead left with $N$ units of R-R flux through the $\mathbf{S}^3$ of a deformed conifold \cite{Klebanov:2000hb}.

Another way to make contact with the ${\cal N}=1$ conifold theory is to start from a more supersymmetric theory, namely a quiver ${\cal N}=2$ field theory \cite{Klebanov:1998hh}. In that case one starts from the orbifold background $\mathbf{S}^5/\mathbb{Z}_2$, that is dual to the ${\cal N}=2$ super-conformal Yang-Mills $U(N)\times U(N)$ theory with hypermultiplets transforming in $(N,\bar{N})\oplus (\bar{N},N)$. From an ${\cal N}=1$ perspective, the hypermultiplets correspond to chiral multiplets and other chiral multiplets in the adjoint representations of the two $U(N)$'s are present. There is also a non-trivial superpotential for the multiplets in the bi-fundamental representations of the gauge group. One can add now a relevant term to this superpotential. To integrate out the adjoint multiplets in presence of this relevant perturbation corresponds to blow up the orbifold singularity of $\mathbf{S}^5/\mathbb{Z}_2$ \cite{Klebanov:1998hh}. Remarkably, in \cite{Klebanov:1998hh} it has also been shown that such blown-up space is topologically equivalent to the coset space $T^{1,1}$, the base of the conifold.

We will describe the flow to the conifold theory in a different way \cite{merlatti}, hoping to contribute to clarify some of its intricate aspects. We start from the ${\cal N}=2$ superconformal Yang-Mills theory. When conformal invariance is broken introducing fractional branes, we find how partial supersymmetry breaking can be described on the supergravity side. Within this purely ten dimensional superstring context, we see then how the resolved conifold background emerges naturally as the proper one to engineer ${\cal N}=1$ super-Yang-Mills theory, with no matter and just one gauge group. We follow all the steps in terms of explicit supergravity solutions.

Having eventually at our disposal a proper gravitational dual description of pure ${\cal N}=1$ super-Yang-Mills theory, many results can be obtained for the latter. Some of them are checks of previously known field theoretical computations. Others are new predictions. The interplay between gauge theory and gravity has indeed a quite rich structure. Among many, we will describe one of the possible applications. We will see how the dual gravitational description can be used to determine the low energy effective superpotential of the field theory under consideration, namely the Veneziano-Yankielowicz \cite{Veneziano:1982ah} superpotential\footnote{In presence of flavors, also the Affleck-Dine-Seiberg \cite{Affleck:1983mk} superpotential can be derived in an analogous way \cite{Imeroni:2003cw}}. The same gravitational description seems also to suggest that other degrees of freedom, with the same quantum number of a glueball, are relevant at low energy. We finally determine the form of the generalized Veneziano-Yankielowicz superpotential, with the inclusion of such glueball states \cite{Merlatti:2004df}. Some physical implications are discussed.

\section{Partial supersymmetry breaking and a new path to the conifold}

We start from the simplest generalization of the original AdS/CFT correspondence, namely we will consider the case in which the internal manifold is not simply $\mathbf{S}^5$, but the orbifold $\mathbf{S}^5/\mathbb{Z}_2$ \cite{Kachru:1998ys}. As discussed in the introduction, the dual gauge theory is ${\cal N}=2$ super-conformal Yang-Mills $U(N)\times U(N)$ theory with hypermultiplets transforming in $(N,\bar{N})\oplus (\bar{N},N)$. As for the original ${\cal N}=4$ case, on the string side we know the explicit supergravity solution connecting the UV geometry (the relevant one for the engineering of the gauge theory) and the IR one. If we parameterize the six dimensional transverse space, ({\it i.e.} the cone over $\mathbf{S}^5/\mathbb{Z}_2$), in terms of the coordinates
\be
X_4 &=& r\sin\xi\cos\tau\\
X_5 &=& r\sin\xi\sin\tau\\
X_6 &=& r\cos\xi\cos\frac{\theta}{2}\cos\frac{\beta+\theta}{2}\\
X_7 &=& r\cos\xi\cos\frac{\theta}{2}\sin\frac{\beta+\theta}{2}\\
X_8 &=& r\cos\xi\sin\frac{\theta}{2}\cos\frac{\beta-\theta}{2}\\
X_9 &=& r\cos\xi\sin\frac{\theta}{2}\sin\frac{\beta-\theta}{2},
\ee
the right prescription for the orbifolding is $0<\beta<2\pi$, $0<\xi<\pi$, $0<\theta<\pi$ and $0<\tau<2\pi$. It is easy to see the $\mathbb{Z}_2$ acts just on the coordinates $X_6,~X_7,~X_8,~X_9$.
The metric of the solitonic solution we are discussing is:
\beq\label{orbi2c}
ds^2=H(r)^{-1/2}dx_{1,3}^2+H(r)^{1/2}\left(dr^2+r^2\Big(\cos^2\xi\sum_{i=1}^3\sigma_i^2+\sin^2\xi d\tau^2+d\xi^2\Big)\right),
\eeq
where the $\sigma_i$ are defined as
\be
2\sigma_1&=&-\sin\beta d\theta+\cos\beta\sin\theta d\phi,\\ \label{sigma}2\sigma_2&=&\cos\beta d\theta+\sin\beta\sin\theta d\phi,\\ 2\sigma_3&=&d\beta+\cos\theta d\phi
\end{eqnarray}
and they satisfy $d\sigma_i=\varepsilon_{ijk}\sigma_j\wedge\sigma_k$.
The explicit expression for $H(r)$ is:
\be
H(r)~=~1+\frac{N}{r^4}.
\ee
We see that for $r\to\infty$ we recover the UV geometry $M_{1,3}$ times the cone over $\mathbf{S}^5/\mathbb{Z}_2$, while for small $r$ the dual $AdS_5\times\mathbf{S}^5/\mathbb{Z}_2$ geometry is found. The $AdS$ factor implies the dual gauge theory is conformal. It is well known that in orbifold spaces  a way to break conformal invariance is to add fractional branes. Such branes are charged under the twisted fields that appear in the orbifold spectrum. As we want to break conformal invariance, we try now to switch on the proper twisted fields (the ones that couple to fractional D3-branes in such background). We start perturbing the same UV geometry and we hope to be able to find a new solution.

Before doing this, we make the following change of coordinates in the six dimensional transverse space:\beq r~=~\sqrt{\rho^2+R^2}\ \ \ ,\ \ \ \tan\xi~=~\frac{\rho}{R}.\eeq
We see that the metric on the cone over $\mathbf{S}^5/\mathbb{Z}_2$ becomes
\beq ds^2_6~=~dR^2+R^2\sum_{i=1}^3\sigma_i^2~+~d\rho^2+\rho^2d\tau^2.\eeq
It is thus easy to see that this space factorizes in the product of a four-dimensional ALE space ($\mathbb{C}^2/\mathbb{Z}_2$, {\it i.e.} the cone over $\mathbf{S}^3/\mathbb{Z}_2$, parameterized by $R,~\beta,~\theta$ and $\phi$) and a two dimensional plane ($\mathbb{R}^2$, parameterized by $\rho$ and $\tau$). In this way we recover the isomorphism between the cone over $\mathbf{S}^5/\mathbb{Z}_2$ and the orbifold space $\mathbb{R}^2\times\mathbb{C}^2/\mathbb{Z}_2$. This isomorphism makes all the picture consistent. The field theory on the $D3$-branes is indeed engineered on the latter orbifold \cite{Douglas:1996sw}. 

We can now disregard the bulk $D3$-branes originating the metric (\ref{orbi2c}) and try to find an analogous solution just for the fractional branes. This has been done explicitely in \cite{Bertolini:2000dk} for the orbifold $\mathbb{R}^2\times\mathbb{C}^2/\mathbb{Z}_2$\footnote{see \cite{billo} for more general orbifolds} and we can borrow their results. They describe  a stack of $N$ fractional D3-branes in the IIB orbifold background (${\mathbb R}^{1,5}\times {\mathbb C}_2/{\mathbf Z}_2$) \cite{Bertolini:2000dk,Polchinski:2000mx}. On their world volume a ${\cal N}=2$ supersymmetric $SU(N)$ Yang-Mills theory lives.

The solution can be written in terms of the metric, a R-R four form potential ($C_4$), a NS-NS twisted scalar ($\tilde{b}$) and a R-R twisted one ($c$):
\begin{eqnarray}\nonumber
ds^2&=&H(r,\rho)^{-1/2}\eta_{\alpha\beta}dx^{\alpha}dx^{\beta}~+~H(r,\rho)^{1/2}\delta_{ij}dx^idx^j,\\\label{orbi}  C_4&=&\left(H(r,\rho)^{-1}-1\right)~dx^0\wedge\ldots\wedge dx^3,\\c&=&~NK~\tau\hspace{1.2cm},\hspace{1.2cm}\tilde{b}~=~NK\log\frac{\rho}{\epsilon}~,\nonumber
\end{eqnarray} where $\alpha,\beta=0,\ldots,3;\ i,j=4,\ldots,9;\ r=\sqrt{\delta_{ij}x^ix^j};\ K=4\pi g_s\alpha';\ \epsilon$ is a regulator; $\rho$ and $\tau$ are polar coordinates in the plane $x^4,\ x^5$. The self-duality constraint on the R-R five-form field strength has to be imposed by hand. 

The precise functional form of the function $H(r,\rho)$ has also been determined \cite{Bertolini:2000dk} but it is not relevant for our purposes. It is enough for us to note that the spacetime (\ref{orbi}) has a naked singularity. We are thus not able to follow our program completely and we are prevented to find the dual geometry in the deep IR. This is a quite general feature of those gravitational backgrounds that are dual to non-conformal gauge theories. Luckily, in many cases the singularities are not actually present but they are removed by various mechanisms. In theories ${\cal N}=2$ supersymmetric, like this one or the similar ``wrapped branes'' example \cite{Gauntlett:2001ps}, the relevant mechanism is the enhan\c{c}on \cite{Johnson:1999qt}. It removes a family of time-like singularities by forming a shell of (massless) branes on which the exterior geometry terminates. As a result, the interior singularities are ``excised'' and the spacetime becomes acceptable, even if the geometry inside the shell has still to be determined. For the case at hand, the appearance of such shell, its field-theoretic interpretation and its consistency at the supergravity level have been discussed in \cite{Bertolini:2000dk,Polchinski:2000mx,Merlatti:2001gd}. It turns out that the enhan\c{c}on shell is simply a ring in the $(\rho,\tau)$-plane. Its radius is \begin{equation}\label{re}\rho_e=\epsilon~ e^{-\pi/(2Ng_s)}.\end{equation} The geometry (\ref{orbi}) is thus valid outside the enhan\c{c}on shell, {\it i. e.} for $\rho\geq\rho_e$. On the gauge theory side the enhan\c{c}on locus corresponds to the locus where the Yang-Mills coupling constant diverges \cite{Bertolini:2000dk,Polchinski:2000mx}.

\subsection{Probe analysis}

To understand better what is going on we can perform a probe analysis\footnote{for a review of this method see, for example, \cite{Johnson:2000ch}} in terms of the fields appearing in (\ref{orbi}) \cite{merlatti}.
We should thus consider the following boundary action
\begin{eqnarray}
\label{boundary1}
S_{bdy}&=&-\frac{T_3}{\sqrt{2}k_{orb}}\int d^4x\sqrt{-det~G_{\alpha\beta}+2\pi\alpha'F_{\alpha\beta}}\left(1+\frac{1}{2\pi^2\alpha'}\tilde{b}\right)\\ \nonumber &&+\frac{T_3}{\sqrt{2}k_{orb}}\int C_4\left(1+\frac{1}{2\pi^2\alpha'}\tilde{b}\right)+\frac{T_3}{\sqrt{2}k_{orb}}\frac{\alpha'}{2}\int c~F\wedge F.
\end{eqnarray}
where $ T_3~=~\sqrt{\pi}$, $k_{orb}~=~(2\pi)^{7/2}g_s\alpha'^2$ and we have also included the world-volume vector fields fluctuations. As usual, the coordinates $x^a=2\pi\alpha'\Phi^a$ are identified with the chiral scalar fields of the Yang-Mills theory living on the world-volume of the brane. Writing now the boundary action (\ref{boundary1}) in terms of the classical solution (\ref{orbi}), we get:
\begin{eqnarray}
 S_{probe}&=&-\frac{T_3}{\sqrt{2}k_{orb}}\Bigg\{\int d^4x\frac{(2\pi\alpha')^2}{2}\left(\frac{1}{2}\delta_{ab}\partial^{\alpha}\Phi^a\partial_{\alpha}\Phi^b+\frac{1}{4}F_a^{\alpha\beta}F^a_{\alpha\beta}\right)\cdot\\ &&\hspace{-1.3cm}\cdot\left(1+~\frac{NK}{2\pi^2\alpha'}\log\frac{\rho}{\epsilon}\right) + V_4\left(1+\frac{NK}{2\pi^2\alpha'}\log\frac{\rho}{\epsilon}\right)-\frac{\alpha'}{2}KN\tau\int F\wedge F\Bigg\}.\nonumber\label{probeaction2}
\end{eqnarray}
We note that a non trivial potential is generated, namely:
\begin{equation}
V(\Phi)~=~\frac{T_3}{\sqrt{2}k_{orb}}\left(1+\frac{NK}{2\pi^2\alpha'}\log\frac{\rho}{\epsilon}\right)~=~\frac{1}{(2\pi)^3\alpha'^2}~ \tau_{Y2}\label{potm}
\end{equation}
where we have defined $\tau_Y\equiv\tau_{Y1}+i\tau_{Y2}\equiv \frac{4\pi i}{g^2_Y}+\frac{\theta_Y}{2\pi}$ as the holographic complexified Yang-Mills coupling \cite{Bertolini:2002xu}:
\begin{equation}\label{varie}\tau_{Y}~=~\frac{1}{(2\pi\sqrt{\alpha'})^2g_s}\left( c+i\left(2\pi^2\alpha'+\tilde{b}\right)\right).\end{equation}

This computation shows that this system violates the no-force condition. This could seem quite surprising as the configuration is still ${\cal N}=2$ supersymmetric and usually the no-force condition is respected with such amount of supersymmetry (on the field theory side such condition reflects the flatness of the Coulomb branch). Nevertheless, at least from a purely field theoretical point of view, we know that such flatness can be violated by electric-magnetic Fayet-Iliopoulos terms \cite{Antoniadis:1995vb}, that preserve ${\cal N}=2$ supersymmetry but generate a non-trivial potential:
\begin{equation}
V~=~\frac{|e~+~m\tau|^2~+~\xi^2}{\tau_2},
\label{pot1}\end{equation}
where $e,~m$ and $\xi$ are free parameters.
Luckily, it turns out the potential (\ref{potm}) is just a specific case of (\ref{pot1}), namely with \begin{equation}e~=~-\frac{m\theta_Y}{2\pi}\ ,\ \ \ \xi=0\ \ \ \mbox{and}\ \ \ m^2~=~\frac{1}{(2\pi)^3\alpha'^2}\label{values}.\end{equation} We see in this way that the violation of the no-force condition does not contradict supersymmetry, but it corresponds to the generation of a purely magnetic FI term. We call it purely magnetic because a dyon with quantum numbers $(e,m)$ has an effective electric charge equal to $e+\frac{m\theta}{2\pi}$ \cite{Witten:1979ey} and with the values in (\ref{values}) it vanishes identically. Such values ensure also that the probe-brane is in a proper vacuum in the $\hat{\tau}$ direction, but for generic values of $\rho$ it is not and it moves toward $\rho=\rho_e$. This situation holographically corresponds to vacua preserving all the supersymmetries but signaling a singularity. This continues to be in perfect agreement with the gravity picture we just got: the probe-brane is pushed to the enhan\c{c}on and there stays. From the gauge theory it is moreover possible to argue \cite{Antoniadis:1995vb} at that point of the moduli space extra massless states appear (i.e. monopoles) and to give a physical interpretation of the singularity. On the gravity side it corresponds to the well known fact that at the enhan\c{c}on extra massless objects appear \cite{Johnson:1999qt} and it enforces the interpretation of the enhan\c{c}on locus as the field theoretic curve of marginal stability \cite{merlatti} (we mean the curve in moduli space where extra massless objects, typically dyons, appear \cite{Ferrari:1996sv}).

\subsection{Away from the orbifold limit} 

We are now going to soften the orbifold singularity and consider the resulting (smooth) Eguchi-Hanson space \cite{Eguchi:1978xp}. We generalize the probe analysis that we have done in the orbifold and find partial supersymmetry breaking. We are then led to consider ${\cal N}=1$ supersymmetric Yang-Mills theory and the resolved conifold geometry \cite{PandoZayas:2000sq}\footnote{see also \cite{Becker:2004qh} for a way to embed the non-supersymmetric solution of \cite{PandoZayas:2000sq} in a supersymmetric set-up}.

The dual field theory analysis of the previous section has been entirely performed in an orbifold background and we have found purely magnetic FI terms. It is well known \cite{Douglas:1996sw} that a way to introduce standard ({\it i.e.} electric) FI terms is via the blow up of the orbifold singularity. In such case the resulting space is no longer an orbifold but it is a (smooth) Eguchi-Hanson ($EH$) space \cite{Eguchi:1978xp}. This is an ALE space. It is characterized by three moduli that correspond to three different ways of blowing-up the relevant two cycle when going away from the orbifold limit (via the triplet of massless NS-NS twisted scalars). Analogously to the field theory description of ${\cal N}=2$ FI terms, it is possible to use the global $SU(2)_R$ symmetry and to reduce to the case of just one modulus. The one-parameter family of metrics we get in this way can be written as:
\begin{equation}\label{EH}
ds^2_{EH}~=~\left( 1-\left(\frac{a}{r}\right)^4\right)^{-1}dr^2~+~r^2\left(1-\left(\frac{a}{r}\right)^4\right)\sigma_3^2~+~r^2(\sigma_1^2+\sigma_2^2),
\end{equation}
where $a\leq r\leq \infty$, $a^2$ is proportional to the size of the blown-up cycle, the $\sigma_i$ are defined in terms of the ${\mathbf S}^3$ Euler angles ($\theta,~\phi,~\beta$) as in (\ref{sigma}).

The solution corresponding to fractional branes in the Eguchi-Hanson background has been found in \cite{Bertolini:2001ma} and it is very similar to the orbifold one. It can be written in terms of a metric
\begin{equation}\label{EHm}
ds^2=~H(r,\rho)^{-1/2}\eta_{\alpha\beta}dx^{\alpha}dx^{\beta}~+~H(r,\rho)^{1/2}\left(d\rho^2+\rho^2d\tau^2\right)~+~H(r,\rho)^{1/2}ds^2_{EH},\end{equation} a self dual R-R five-form\begin{equation}F_5~=~d\left(H(r,\rho)^{-1}~dx^0\wedge\ldots\wedge dx^3\right)~+~\star d\left(H(r,\rho)^{-1}~dx^0\wedge\ldots\wedge dx^3\right),\end{equation} and a complex three form (a linear combination of the NS-NS one and the R-R one) valued only in the transverse space:\begin{equation}\label{EH3} {\mathcal H}~\equiv~F_3^{NS}+iF_3^{R} ~=~d\gamma(\rho,\tau)\wedge\omega,\end{equation}
where $\gamma(\rho,\tau)$ is an analytic function on the transverse $\mathbb{R}^2$, $\omega$ is the harmonic anti-self-dual form on the Eguchi-Hanson space and $r$ is now the radial Eguchi-Hanson coordinate.

Even if the microscopic interpretation of this solution is quite obscure \cite{Bertolini:2001ma}, we assume that it corresponds to the standard geometrical interpretation of fractional $D3$-branes as $D5$-branes wrapped on the non-trivial cycle that shrinks to zero size in the orbifold limit \cite{Douglas:1996xg}. The form of the solution is indeed consistent with the interpretation of twisted fields as fields coming from wrapping of higher degree forms \cite{Douglas:1996xg}, as it was already confirmed in \cite{Merlatti:2000ne} by explicit computations of string scattering amplitudes. The main difference with respect to the orbifold solution is the explicit functional form of the warp factor $H(r,\rho)$, whose physical implication is that there is still a naked singularity screened by an enhan\c{c}on mechanism, but contrary to the orbifold case, in the internal directions (those along the ALE space) the singularity is cured by the blowing-up of the cycle and there is no enhan\c{c}on there \cite{Bertolini:2001ma}.

Unluckily, the Eguchi-Hanson background lacks of a conformal field theory description. This makes it difficult to determine the correct form of the boundary actions one would need to get gauge theory information. It looks nevertheless quite reasonable to assume the proper boundary action is the same as the orbifold one with the inclusion of the new FI term corresponding to having blown up the cycle. This can be motivated by the knowledge of the explicit solution in the Eguchi-Hanson case, that makes it clear the close similarity to its orbifold limit. Besides, this is very analogous in spirit to the successful assumption made in \cite{Polchinski:1996ry}. The potential the brane fills will be thus the same we computed in the orbifold (we refer to the magnetic case (\ref{potm})) but now with an extra term corresponding to have $\xi\neq 0$ and proportional to the size of the blown-up cycle ($a^2$):
\begin{equation}
V(\Phi)~=~m^2 \tau_2+\frac{\xi^2}{\tau_2}\label{potem}
\end{equation}
We see again that the no-force condition is violated but now there is an interesting minimum where the probe-brane can sit, i.e. at \begin{equation}\rho_*~=~\rho_e e^{\frac{\pi}{N}\frac{\xi}{m}},\label{minimum*}\end{equation}where the enhan\c{c}on radius $\rho_e$ has been defined in (\ref{re}). As $\rho_*>\rho_e$, $\rho=\rho_*$ defines a regular space-time locus. From the field theory analysis \cite{Antoniadis:1995vb} we know this kind of vacuum preserves just half of the supersymmetry. This implies that \cite{merlatti} the gauge theory living on the probe is ${\cal N}=1$ supersymmetric with a massless vector multiplet and a massive chiral one, with mass $M$ given by:\begin{equation}\label{mass}M=\frac{m^2}{2\xi}\langle\tau'\rangle.\end{equation} From the supergravity point of view we have again a solution with bulk ${\cal N}=2$ supersymmetry (as shown in \cite{Bertolini:2001ma}) but the embedding of the brane probe seems to break half of it. It would be interesting to see this purely in a stringy context by making a proper K-supersymmetry study of this geometry. However, our field theory analysis already shows that the geometry seen by the probe-brane is locally ${\cal N}=1$ supersymmetric. 

A careful analysis actually shows \cite{merlatti} the geometry the probe-brane sees is locally the same as the ``near horizon'' of the resolved conifold one.  
It is not surprising we find the resolved conifold geometry appearing here: it is indeed known the resolved conifold geometry is the proper one to engineer ${\cal N}=1$ pure supersymmetric Yang-Mills theory in the type IIB set-up \cite{Vafa:2000wi}. In this talk we are just following this engineering entirely in terms of supergravity solutions and we clearly see its relation with the parent ${\cal N}=2$ orbifold theory. A non-trivial information we immediately gain (with the help of the field theory analysis and particularly thanks to the formula (\ref{mass})) is that the size of the blown-up two-cycle is directly related to the inverse mass of the chiral multiplet. 

The assumption about the microscopic interpretation of the probe as wrapped $D5$-branes cannot be proven here, but, as we will see, it makes all the picture consistent. Under this assumption, we have learned that, if a $D5$-brane is wrapped on the non-trivial two-cycle of a resolved conifold, close to the apex of such cone the gauge theory living on its world-volume is ${\cal N}=1$ supersymmetric. The embedding of $N$ branes in such background will then give rise to ${\cal N}=1$ supersymmetric $U(N)$ Yang-Mills theory. The fact we reach such conclusion via a probe analysis means that we are studying open strings (and then branes) in the given background. This matches perfectly with the picture of geometric transition we discussed in the introduction: being talking of open strings and branes, we are before the transition and accordingly the background is the resolved conifold! 

To get pure ${\cal N}=1$ super-Yang-Mills, we take the zero size limit of the blown-up cycle. In this way the chiral adjoint field acquires infinite mass (see eq. (\ref{mass})) and it is decoupled. The relevant geometry is now the conifold one. This geometry is still singular \cite{Klebanov:2000nc} and according to equation (\ref{minimum*}) the brane will tend to go to the enhan\c{c}on locus. Consistently, it has been shown that at least from a supergravity point of view the enhan\c{c}on is a consistent mechanism also in the conifold case \cite{Merlatti:2001gd}. From \cite{Vafa:2000wi,Klebanov:2000hb}, we know how the story goes on, but we briefly review it here following the perspective we have just developed. In principle there are two ways now to de-singularize the geometry (or we could also say to determine the geometry inside the enhan\c{c}on shell): the $\mathbf{S}^2$ or the $\mathbf{S}^3$ can be blown-up. As we don't want to go back to the original case, the only remaining possibility is to look for a consistent deformed conifold solution. From \cite{Klebanov:2000hb} we know this is indeed the right thing to do: the deformed conifold is a non-singular solution and its geometry is the proper one to describe the infrared ${\cal N}=1$ Yang-Mills dynamics.

\section{Effective superpotentials from supergravity}

In this section we briefly review the derivation of gauge theory superpotentials from proper gravitational duals. 

The best way to do it is to follow the picture of geometric transition \cite{Vafa:2000wi,Cachazo:2001jy} we described in the introduction. In this set-up,
from the knowledge of the geometry after the transition and the map between the original microscopic field theoretical degrees of freedom and the new geometrical data, it is indeed possible to determine the effective gauge theory superpotential. 
We now briefly see how it does work for the ${\cal N}=1$ conifold geometry. Quite generally, it is possible to write the supergravity solution in terms of a superpotential \cite{Taylor:1999ii}, namely:
\begin{equation}
W_{eff}=\int_AG_3\int_B\Omega~-~\int_A\Omega\int_BG_3
,\end{equation}
where $G_3=F_3+\tau H_3$, $\tau=C_0+i {\mbox e}^{-\Phi}$ and
$\Omega$ is the holomorphic $(3,0)$ form of the Calabi-Yau (our conifold).
$A$ and $B$ are proper 3-cycles ($A$ is the compact one, while $B$ is non compact).

To translate this superpotential in a field theory one, we need a map relating geometric quantities to gauge parameters. According to the proposal in \cite{Cachazo:2001jy}, based on the topological string in the way we sketched in the introduction, this map is: 
\begin{equation}\label{ggmap}
\int_AG_3=N \hspace{0.3cm},\hspace{0.3cm}\int_BG_3=\frac{1}{2\pi i}~\frac{8\pi^2}{g_{YM}^2}\hspace{0.3cm},\hspace{0.3cm}\int_A\Omega=2\pi i~ S\end{equation}where
\begin{equation}\nonumber S~=~\frac{1}{16\pi^2} Tr W^2~=~-\frac{1}{32\pi^2}\lambda^{\alpha}\lambda_{\alpha}+\ldots\end{equation}

Following this proposal, it is now possible to determine the effective gauge theory superpotential \cite{Cachazo:2001jy}. It turns out to be:
\begin{equation}
W_{V.Y.}(S,\Lambda_0,g^2_{YM})=NS\left(\ln\frac{S}{\Lambda_0^3}+\frac{8\pi^2}{N g_{YM}^2}-1\right)
\label{vy}\end{equation}
where constant terms and subleading powers of $\Lambda_0$ (the ultra-violet cut-off at which $g_{YM}$ is evaluated) have been neglected. 
From the knowledge of the one loop $\beta$-function it is easy to see this is precisely the Veneziano-Yankielowicz superpotential \cite{Veneziano:1982ah}.

\subsection{An extension of the Veneziano-Yankielowicz superpotential}

As it has been pointed out in \cite{Imeroni:2003jk}, it is also possible to find relations similar to (\ref{ggmap}) from supergravity alone and some basic geometric considerations. In that case the formula concerning the fluxes in (\ref{ggmap}) can be written as:
\begin{equation}\label{ggnew}
\int_A F_3~=~N,\hspace{.9cm}\int_B F_3~=~0,\hspace{.9cm}\int_B H_3~=~-\frac{4\pi}{g_{YM}^2},\hspace{.9cm}\int_A H_3~=~0,
\end{equation}
where for simplicity we are considering the $\theta_{YM}=0$ sector of the field theory.
 Of course the two formulas (\ref{ggmap}) and (\ref{ggnew}) are completely equivalent in the case considered here, being $\langle\tau\rangle=i$. The next step we want to do is to determine the form of the gauge-theory superpotential when going off-shell for the $\tau$-field, namely if we try to include its oscillations ($\tau~=~\langle\tau\rangle~+~i~\delta$). To this aim, formulas (\ref{ggnew}) are needed and we easily get:
\begin{equation}
W_{eff}(S,\delta)=NS\left(\ln\frac{S}{\Lambda^3}+\frac{8\pi^2}{N g_{YM}^2}\delta-1\right)
\end{equation}
where we have already used the one-loop $\beta$-function to relate $\Lambda_0,\ g_{YM}$ and $\Lambda$.

Now we would like to interpret the new modulus $\delta$ from the gauge theory point of view. This turns out to be very easy as the shift of the dilaton field is known to be related to the glueball field $0^{++}$ \cite{Witten:1998qj,Gubser:1998bc}. This fact has been used in \cite{Caceres:2000qe} to compute the $0^{++}$ mass. The RR scalar is instead related to the glueball $0^{+-}$. Thus all in all, the complex field $\delta$ is related to the complex glueball field. Even if we don't know the precise formula of such a relation, we can generically write:
\begin{equation}
\frac{8\pi^2}{g^2_{YM}}\delta~=~f(\chi)
\end{equation}
where $\chi$ is now the glueball superfield and $f(\chi)$ is an undetermined function of it. We get finally for the gauge theory superpotential the expression: \begin{equation}\label{MSVY}W(S,\chi)=NS\left(\ln\frac{S}{\Lambda^3}-1\right) +S f(\chi)\end{equation}

It is quite amazing that this superpotential is the same as the extended Veneziano-Yankielowicz superpotential determined in \cite{Merlatti:2004df}. Also there the new $\chi$ degrees of freedom were interpreted as glueball superfield.
 A crucial difference between the two approaches (the one presented here and the one in \cite{Merlatti:2004df}) is that the supergravity solution we consider here already sits at a precise vacuum of the gaugino condensate composite superfield and consequently the superpotential one can extract has also just one minimum and does not reproduce the complete vacuum structure of the model. This is why in this case we get $f(\langle \chi\rangle)=0$ instead of the more general result $f(\langle\chi\rangle)=2\pi i k$ as in \cite{Merlatti:2004df}. The inclusion of this degree of freedom could help also in finding a proper dual description of domain walls, as it has been suggested in \cite{Merlatti:2005sd} for the field theory.

However, we have just seen that the gauge/gravity correspondence suggests an extension of the Veneziano-Yankielowicz superpotential so to include also glueball degrees of freedom. The general form of such extended superpotential is (\ref{MSVY}), with $f(\chi)$ an holomorphic function of $\chi$.
The determination of the function $f(\chi)$ would now shed light on
the super Yang-Mills infrared properties. In \cite{Merlatti:2004df} it has been shown that various arguments  point to the same function:
\begin{eqnarray}\label{effe}
f(\chi) = N\ln \left[-e\frac{\chi}{N}\ln \chi^N\right] \ .
\end{eqnarray}
This function passes a number of consistency checks \cite{Merlatti:2004df}: i)
the Veneziano-Yankielowicz superpotential is recovered when the glueball superfield is integrated
out. Besides this procedure naturally leads to the $N$ independent
vacua of the theory. ii) Non
supersymmetric gluodynamics is better approached when giving a mass to 
the gluino. The
theory leads to a potential which resembles the ordinary glueball
effective potential for the Yang-Mills theory. iii) A reasonable
integrating in method leads to the same function.
Besides, in \cite{Feo:2004mr} such extended superpotential has been used to compute the mass spectrum of the lightest states of ${\cal N}=1$ super Yang-Mills. It has been concluded that the lightest state is the gluinoball field and it has a small mixing with the glueball state.

\section{Conclusions}

In this talk we have described how partial supersymmetry breaking can be obtained in the context of the gauge/string correspondence. This has been reached studying the low energy dynamics of open strings in given backgrounds. The tool we used to perform such study is the probe-analysis. We have seen in this way that a potential, due to the internal flux, is generated in the ${\cal N}=2$ orbifold. This potential drives the probe-brane to the enhan\c{c}on locus. Instead, when the orbifold singularity is softened to the smooth Eguchi-Hanson space, the resulting potential drives the probe-brane to regular space-time points, ({\it i.e.} outside the enhan\c{c}on shell). We have found that at those points (corresponding to minima of the potential) partial supersymmetry breaking does occur. We have further analyzed the geometry seen by the probe-brane close to those minima and we have found it is the resolved conifold one, in the vicinity of the blown-up two-cycle. We have finally showed that shrinking the two-cycle to zero size corresponds, on the field theory side, to decouple the adjoint chiral scalar and thus to get pure ${\cal N}=1$ super-Yang-Mills. 

We have then used the gravitational background dual to such theory to determine the low energy effective superpotential. Besides the standard Veneziano-Yankielowicz result we have seen that glueball can indeed be included in this picture. We have then discussed the physical implications of the inclusion of this new degree of freedom at low energy. We have found it is heavier than the gluinoball.

\vskip 1cm \centerline{\bf Acknowledgments} \noindent I thank A. Lerda for various discussions and exchange of ideas. This work is partly supported by MCyT and FEDER under grant FPA2005-00188, by Xunta de Galicia (grant PGIDIT06PXIB206185PR and Conselleria de Educacion) and by the EC Commision under grant MRTN-CT-2004-005104.



\begin{thebibliography}{99} 
 
\bibitem{Maldacena:1997re}
J.~M.~Maldacena,
Adv.\ Theor.\ Math.\ Phys.\  {\bf 2} (1998) 231
[Int.\ J.\ Theor.\ Phys.\  {\bf 38} (1999) 1113]

\bibitem{Aharony:1999ti}
O.~Aharony, S.~S.~Gubser, J.~M.~Maldacena, H.~Ooguri and Y.~Oz,
Phys.\ Rept.\  {\bf 323} (2000) 183

\bibitem{Aharony:2002up}
O.~Aharony,
arXiv:hep-th/0212193.

M.~Bertolini,
Int.\ J.\ Mod.\ Phys.\ A {\bf 18} (2003) 5647 

F.~Bigazzi, A.~L.~Cotrone, M.~Petrini and A.~Zaffaroni,
Riv.\ Nuovo Cim.\  {\bf 25N12} (2002) 1

P.~Di Vecchia, A.~Liccardo, R.~Marotta and F.~Pezzella,
Int.\ J.\ Mod.\ Phys.\ A {\bf 20} (2005) 4699

J.~D.~Edelstein and R.~Portugues,
arXiv:hep-th/0602021.

\bibitem{Imeroni:2003jk}
E.~Imeroni,
arXiv:hep-th/0312070.



\bibitem{Maldacena:2000yy}
J.~M.~Maldacena and C.~Nunez,
Phys.\ Rev.\ Lett.\  {\bf 86} (2001) 588

\bibitem{Klebanov:2000hb} I.R. Klebanov and M.J. Strassler, 
JHEP {\bf 08} (2000) 
052, 





\bibitem{Gopakumar:1998ki}
R.~Gopakumar and C.~Vafa,
Adv.\ Theor.\ Math.\ Phys.\  {\bf 3} (1999) 1415

\bibitem{Witten:1992fb}
E.~Witten,
Prog.\ Math.\  {\bf 133} (1995) 637


\bibitem{Vafa:2000wi}
C.~Vafa,
J.\ Math.\ Phys.\  {\bf 42} (2001) 2798


\bibitem{Cachazo:2001jy}
F.~Cachazo, K.~A.~Intriligator and C.~Vafa,
Nucl.\ Phys.\ B {\bf 603} (2001) 3


\bibitem{Klebanov:1998hh}
I.~R.~Klebanov and E.~Witten,
Nucl.\ Phys.\ B {\bf 536} (1998) 199



\bibitem{merlatti}
P.~Merlatti,
Nucl.\ Phys.\ B {\bf 744} (2006) 207

\bibitem{Veneziano:1982ah}
G.~Veneziano and S.~Yankielowicz,
Phys.\ Lett.\ B {\bf 113} (1982) 231.

\bibitem{Affleck:1983mk}
I.~Affleck, M.~Dine and N.~Seiberg,
Nucl.\ Phys.\ B {\bf 241} (1984) 493.

\bibitem{Imeroni:2003cw}
E.~Imeroni and A.~Lerda,
JHEP {\bf 0312} (2003) 051

\bibitem{Merlatti:2004df}
P.~Merlatti and F.~Sannino,
Phys.\ Rev.\ D {\bf 70} (2004) 065022

\bibitem{Kachru:1998ys}
S.~Kachru and E.~Silverstein,
Phys.\ Rev.\ Lett.\  {\bf 80} (1998) 4855

\bibitem{Douglas:1996sw}
M.~R.~Douglas and G.~W.~Moore,
arXiv:hep-th/9603167.



\bibitem{Bertolini:2000dk}
M.~Bertolini, P.~Di Vecchia, M.~Frau, A.~Lerda, R.~Marotta and I.~Pesando,
JHEP {\bf 0102} (2001) 014

\bibitem{billo}
M. Billo, L.~Gallot and A.~Liccardo,
Nucl.\ Phys.\ B {\bf 614} (2001) 254


\bibitem{Polchinski:2000mx}
J.~Polchinski,
Int.\ J.\ Mod.\ Phys.\ A {\bf 16} (2001) 707

\bibitem{Gauntlett:2001ps}
J.~P.~Gauntlett, N.~Kim, D.~Martelli and D.~Waldram,
Phys.\ Rev.\ D {\bf 64} (2001) 106008

F.~Bigazzi, A.~L.~Cotrone and A.~Zaffaroni,
Phys.\ Lett.\ B {\bf 519} (2001) 269

P.~Di Vecchia, A.~Lerda and P.~Merlatti,
Nucl.\ Phys.\ B {\bf 646} (2002) 43

\bibitem{Johnson:1999qt}
C.~V.~Johnson, A.~W.~Peet and J.~Polchinski,
Phys.\ Rev.\ D {\bf 61} (2000) 086001



\bibitem{Merlatti:2001gd}
P.~Merlatti,
Nucl.\ Phys.\ B {\bf 624} (2002) 200

\bibitem{Johnson:2000ch}
C.~V.~Johnson,
arXiv:hep-th/0007170.

\bibitem{Bertolini:2002xu}
M.~Bertolini, P.~Di Vecchia, M.~Frau, A.~Lerda and R.~Marotta,
Phys.\ Lett.\ B {\bf 540} (2002) 104


\bibitem{Antoniadis:1995vb}
I.~Antoniadis, H.~Partouche and T.~R.~Taylor,
Phys.\ Lett.\ B {\bf 372} (1996) 83

H.~Partouche and B.~Pioline,
Nucl.\ Phys.\ Proc.\ Suppl.\  {\bf 56B} (1997) 322

\bibitem{Witten:1979ey}
E.~Witten,
Phys.\ Lett.\ B {\bf 86}, 283 (1979).

\bibitem{Ferrari:1996sv}
A.~Fayyazuddin,
Mod.\ Phys.\ Lett.\ A {\bf 10} (1995) 2703

F.~Ferrari and A.~Bilal,
Nucl.\ Phys.\ B {\bf 469} (1996) 387


\bibitem{Eguchi:1978xp}
T.~Eguchi and A.~J.~Hanson,
Phys.\ Lett.\ B {\bf 74} (1978) 249.

\bibitem{PandoZayas:2000sq}
L.~A.~Pando Zayas and A.~A.~Tseytlin,
JHEP {\bf 0011} (2000) 028

\bibitem{Becker:2004qh}
M.~Becker, K.~Dasgupta, A.~Knauf and R.~Tatar,
Nucl.\ Phys.\ B {\bf 702} (2004) 207

S.~Alexander, K.~Becker, M.~Becker, K.~Dasgupta, A.~Knauf and R.~Tatar,
Nucl.\ Phys.\ B {\bf 704} (2005) 231

M.~Becker, K.~Dasgupta, S.~Katz, A.~Knauf and R.~Tatar,
arXiv:hep-th/0511099.


\bibitem{Bertolini:2001ma}
M.~Bertolini, V.~L.~Campos, G.~Ferretti, P.~Fre', P.~Salomonson and M.~Trigiante,
Nucl.\ Phys.\ B {\bf 617} (2001) 3

\bibitem{Douglas:1996xg}
M.~R.~Douglas,
JHEP {\bf 9707} (1997) 004

D.~E.~Diaconescu, M.~R.~Douglas and J.~Gomis,
JHEP {\bf 9802} (1998) 013


\bibitem{Merlatti:2000ne}
P.~Merlatti and G.~Sabella,
Nucl.\ Phys.\ B {\bf 602} (2001) 453


\bibitem{Polchinski:1996ry}
J.~Polchinski,
Phys.\ Rev.\ D {\bf 55} (1997) 6423

\bibitem{Klebanov:2000nc}
I.~R.~Klebanov and A.~A.~Tseytlin,
Nucl.\ Phys.\ B {\bf 578} (2000) 123






























\bibitem{Taylor:1999ii}
T.~R.~Taylor and C.~Vafa,
Phys.\ Lett.\ B {\bf 474} (2000) 130


\bibitem{Witten:1998qj}
E.~Witten,
Adv.\ Theor.\ Math.\ Phys.\  {\bf 2} (1998) 253

\bibitem{Gubser:1998bc}
S.~S.~Gubser, I.~R.~Klebanov and A.~M.~Polyakov,
Phys.\ Lett.\ B {\bf 428} (1998) 105

\bibitem{Caceres:2000qe}
E.~Caceres and R.~Hernandez,
Phys.\ Lett.\ B {\bf 504} (2001) 64

\bibitem{Merlatti:2005sd}
P.~Merlatti, F.~Sannino, G.~Vallone and F.~Vian,
Phys.\ Rev.\ D {\bf 71} (2005) 125014

\bibitem{Feo:2004mr}
A.~Feo, P.~Merlatti and F.~Sannino,
Phys.\ Rev.\ D {\bf 70} (2004) 096004














\end{thebibliography}
\end{document}